# Topological valleytronics: Brought to light

*The topological valley Hall effect was predicted as a consequence of the bulk topology of electronic systems. Now it has been observed in photonic crystals, showing that both topology and valley are innate to classical as well as quantum systems.*

Fan Zhang [*e-mail: zhang@utdallas.edu*]

*Department of Physics, University of Texas at Dallas, Richardson, TX 75080, USA*

Valleys — the degenerate energy extrema in the band structure of a material — have recently emerged as novel carriers of information and energy, much like charge and spin. And just like electrons with opposite spin, electrons in opposite valleys can counter-propagate along the edge of a two-dimensional insulator without any dissipation [1, 2]. These symmetry-protected topological effects are usually associated with quantum systems, but classical waves can possess a valley degree of freedom, too. So even though they can carry neither charge nor spin, classical systems should be capable of hosting such exotic topological effects. Now, writing in *Nature Physics*, Fei Gao and coworkers [3] report the unambiguous observation of topological valley transport of electromagnetic waves along a domain wall in a honeycomb photonic crystal.

In 1879, Edwin Hall discovered that when an electric current flows through a thin, flat conductor under a perpendicular magnetic field, the moving charge carriers are deflected to one side of the conductor by the Lorentz force (Fig. 1a). Fundamentally, it is the breaking of time-reversal symmetry that leads to this Hall effect. In 2004, Charles Kane and Eugene Mele predicted a quantum version of a related effect (Fig. 1d): a topological insulator [1] that has an odd number of pairs of counter-propagating edge states with opposite spin (Fig. 1e). In this case, the time-reversal symmetry is unbroken and plays an essential role in protecting this quantum spin Hall effect. Now, the idea of utilizing the combination of symmetry and topology has also inspired the design of novel functionalities in photonic, mechanical, and sonic systems [4].

Analogous to the quantum spin Hall effect, topological valley Hall effects were proposed [2, 5-8] for the situation where the valley is a good quantum number (Fig. 1f). In the quantum spin Hall effect, the time-reversal symmetry does not require spin conservation. By sharp contrast, the valley needs to be a good quantum number in order to validate the bulk topological invariant — the valley Chern number — which corresponds to the number of counter-propagating, valley-projected, chiral edge channels (Fig. 1g). Therefore, inter-valley or lattice-scale scattering needs to be negligibly weak both in the bulk and at the edge [7]. This poses a serious challenge in electronic experiments where often the edge termination or reconstruction will strongly mix the two opposite valleys.

For bilayer graphene with a band gap induced by an external electric field, very smooth domain walls (Fig. 1h) that either switch the layer stacking orders [7-9] or reverse the electric field orientation [5-7, 10] have been used to produce a change of the bulk valley Chern number while minimizing the inter-valley edge backscattering. Indeed, metallic mid-gap states have been experimentally observed along such domain walls, and their conductance has appeared to approach two conductance quanta [9, 10]. However, its full quantization is still under debate. For similar systems without the smooth domain walls, only considerable non-local resistance that scales cubically with local resistance have been detected [11-13]. It is most likely that the mid-gap edge states have become gapped and passivated toward the bulk band edges due to the inevitable inter-valley edge backscattering.

The work done by Gao *et al.* represents an important step in realizing valley Hall effects [2], provides evidence for their topological origin, and thereby settles the debate in the electronics community [5-13]. For both transverse electric and transverse magnetic modes, the valley Chern number changes by one across the valley-preserving zigzag domain wall in their photonic crystal. Indeed, they observed that there is one pair of edge states counter-propagating in the two opposite valleys for each mode and that they can be easily tuned and selectively excited. By refracting these valley-projected chiral modes into free space, they showed that the reflectance is, on average, less than 0.1% across the entire band gap and independent of the detailed shape of domain wall.

Due to the high efficiency of the coupling between the topological modes and free-space modes, it is tempting to envisage practical valleytronic applications in directional antennas, lasers and other communication devices across the electromagnetic spectrum. More fundamentally, the in-phase and out-of-phase relations between the transverse magnetic and transverse electric modes can be exploited to emulate the spin degree of freedom and then introduce an analogue of spin-orbit couplings for electromagnetic waves. This would greatly enrich the topological features [2] of the photonic honeycomb crystal [3] and possibly enable distinct classes of topological photonics.

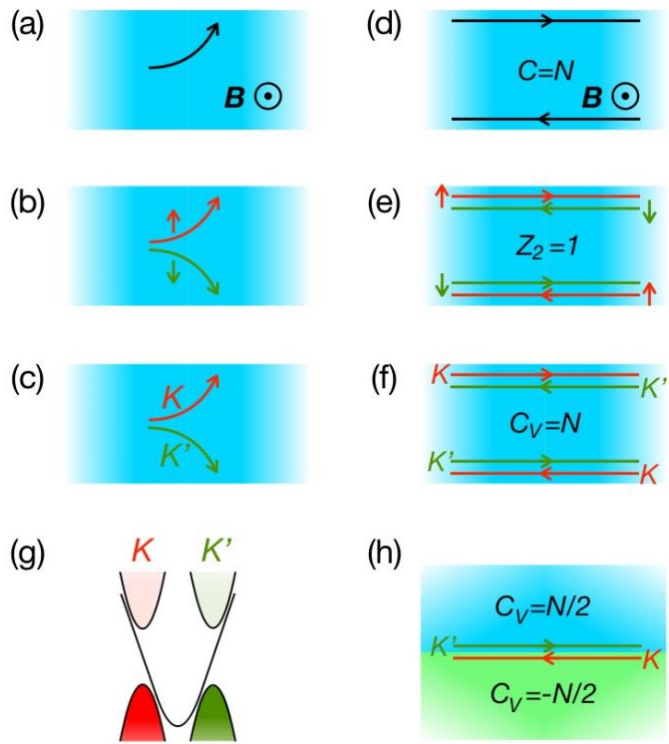

**Figure 1|** Geometric and topological Hall effects. **a-c**, Schematics of the charge, spin, and valley Hall effects in the bulk of 2D systems. **d-f**, the topological counterparts of **a-c**; each displays a band gap in the 2D bulk and protected gapless states at the 1D edge, as a consequence of the nontrivial bulk topological invariant. Respectively, $C$, $Z_2$, and $C_V$ are the first Chern number, the Kane-Mele index [1] and the valley Chern number [2]. **g**, Schematic band structure of a system that exhibits the topological valley Hall effect in **h**. **h**, The boundary between crystal ($C_V=N$) and vacuum ($C_V=0$) in **f** replaced by a smooth domain wall between two topologically distinct crystals ($C_V=\pm N/2$) to reduce the inter-valley edge backscattering that may passivate the counter-propagating valley-projected chiral edge states. We have chosen $N=1$ to illustrate the edge states in **d** and **f-h**.

Together with other works in electronic and sonic systems [4, 9-13], the work of Gao *et al.* suggests that the topological valley Hall effect is universal to both quantum and classical waves. Various valley-projected edge states originating from band topology [2] exist not only in quantum solids but also in artificial crystals. Amazingly, the equatorial ocean and atmospheric waves share a similar topological origin [14]. As a final remark, like the spin Hall effect (Fig. 1b), the valley Hall effect can also be geometric [15], distinct from the topological phenomena [1-14] discussed here. In the geometric valley Hall effects (Fig. 1c), bulk wavepackets such as excitons [16] can bifurcate due to the presence of valley-dependent Berry curvature.

The discovery of counterflow in opposite valleys not only demonstrates that the valley can be a novel carrier of information and energy — particularly valuable for systems without charge and spin — but also exemplifies that symmetry-protected band topology [1, 2] can be universal to both quantum and classical systems. This result is truly elegant and inspiring.